\documentclass[preprint,showpacs,preprintnumbers,amsmath,amssymb]{revtex4}

\usepackage{graphicx}
\usepackage{dcolumn}
\usepackage{bm}

\begin{document}

\noindent

\preprint{}

\title{Statistical deviation from stationary action to Schr\"odinger equation}

\author{Agung Budiyono}

\affiliation{Institute for the Physical and Chemical Research, RIKEN, 2-1 Hirosawa, Wako-shi, Saitama 351-0198, Japan}

\date{\today}

\begin{abstract} 

We discuss the dynamics of single particle by laying a hypothesis that the Hamilton's principle of stationary action is not exact. We then postulate that the deviation of the action with sufficiently short time interval from the stationary action is distributed along a sufficiently long trajectory according to an exponential law. We show that the dynamics of the ensemble of trajectories satisfies the Schr\"odinger equation with Born interpretation of wave function if the average deviation is given by $\hbar/2$ and if two opposite signs of deviation occur equally probably. The particle thus behaves as if it is guided by a virtual wave satisfying the Schr\"odinger equation.    

\end{abstract}

\maketitle 

Let us discuss the dynamics of a particle with mass $m$ in one spatial dimension $x$ subjected to a potential $V(x,t)$, where $t$ denotes time. To do this, let us define action along a segment of path connecting two spacetime points as follows
\begin{equation}
S=\int pdx-Hdt,
\label{infinitesimal action}
\end{equation} 
where $p=mv=\partial_xS$ and $H=-\partial_tS$ are the momentum and energy of the particle, respectively. Here $v=dx/dt$ is the velocity of the particle and we assume that $H$ depends on $x$ and $p$. The Hamilton's principle then says that the only admissible trajectory that connects the two points is given by the one in which $\Delta S$ is stationary \cite{Landau book}. Mathematically, it is then expressed as 
\begin{equation}
\delta S_c=0,
\label{stationary action principle}
\end{equation}
where the variation is done by keeping the end points fixed, and we have denoted the stationary action as $S_c(x,t)$. All the other paths are classically forbidden.

Performing the variation of Eq. (\ref{stationary action principle}), one obtains the Hamilton equation  $dp_c/dt=-\partial_x H_c$ and $v_c=dx/dt=\partial_{p_c} H_c$ \cite{Landau book}, where we have denoted the classical momentum, velocity and energy as $p_c$, $v_c$ and $H_c$. Given a specific dynamical problem, the dynamical equation is thus determined by choosing a specific form of classical energy $H_c$ as function of $p_c$ and $x$. In particular, the Hamilton equation is equivalent to Newton equation if we choose the classical Hamiltonian $H_c$ as follows: 
\begin{equation}
-\partial_tS_c=H_c=\frac{p_c^2}{2m}+V=\frac{(\partial_xS_c)^2}{2m}+V. 
\label{Hamilton-Jacobi equation}
\end{equation}
Viewed as partial differential equation for $S_c(x,t)$, Eq. (\ref{Hamilton-Jacobi equation}) is also called as Hamilton-Jacobi equation. 

To solve the Hamilton-Jacobi equation one needs to choose the initial stationary action $S_c(x,0)$ which implies an initial classical momentum field $p_c(x,0)=\partial_xS_c$. A single trajectory is then picked up if one fixes the initial position as well. If one considers ensemble of initial position with distribution $\rho(x,0)$, then the distribution of the position of the particle at any time $t$ is obtained by solving the continuity equation 
\begin{equation}
\partial_t\rho=-\partial_x(\rho v_c)=-\partial_x\Big(\rho\frac{\partial_xS_c}{m}\Big).  
\label{classical continuity equation}
\end{equation}
Equation (\ref{classical continuity equation}) conserves the local probability flow. 

Let us now exercise an assumption that the Hamilton's principle is not valid exactly but retain the deterministic character of the dynamics. Further let us assume that the short segment trajectory with stationary action is not the only admissible segment of trajectory but the most probable one. This assumption tells us that if we look at a trajectory within a large time scale and average short time fluctuations around the stationary action, then the trajectory looks effectively classical. However, probing the dynamics within sufficiently short time scale will reveal that the momentum and energy are fluctuating around the corresponding classical values. 

To proceed one needs to know how the deviation from the stationary action is distributed along the trajectory. Let us consider a trajectory running from $t=0$ to $t=T$, slice it into sufficiently short segments of equal time interval $\Delta t=T/N$ where $N$ is the number of segments, and take statistics on the distribution of deviation from stationary action $|\Delta S-\Delta S_c|$ along each segment. Let us then postulate that there is a very short universal time interval $\Delta t$ such that using $\Delta t$ to slice a sufficiently long trajectory so that $N$ is sufficiently large, the distribution of $|\Delta S-\Delta S_c|$ converges into exponential law as \cite{book on probability}
\begin{eqnarray}
\mathcal{P}(|\Delta S-\Delta S_c|)\equiv\lim_{N\rightarrow\infty}\frac{n(|\Delta S-\Delta S_c|)}{N}\nonumber\\\sim\exp\Big\{-\frac{2}{\hbar}\Big|\Delta S-\Delta S_c\Big|\Big\},
\label{postulate of exponential distribution}
\end{eqnarray}
where $n(|\Delta S-\Delta S_c|)$ is the number of short segments whose deviation from the stationary action is $|\Delta S-\Delta S_c|$, and $\hbar$ is a constant with action dimension. The average deviation is thus $\hbar/2$. Note that Eq. (\ref{postulate of exponential distribution}) should not be interpreted to give the probability of an elementary step of a stochastic dynamics.

Since our basic law takes statistical form then it is impossible to develop dynamical causal relation which refers to single event (trajectory) as in classical mechanics. Hence, one is forced to instead consider an ensemble of trajectories described by $\rho(x,t)$ which is transported along the deterministic momentum flow $p(x,t)=\partial_xS$. Now, let us consider $\rho$ at two very close spacetime points $\{x,t\}$ and $\{x+\Delta x,t+\Delta t\}$ connected by a short segment of trajectory with an action $\Delta S$. Since the flow is deterministic and since we are only given the distribution of deviation from the stationary action $|\Delta S-\Delta S_c|$ along the trajectory, then $\rho(x+\Delta x,t+\Delta t)$ has to be proportional to $\rho(x,t)$ multiplied by the chance that the short segment with deviation $|\Delta S-\Delta S_c|$ occurs which is given by Eq. (\ref{postulate of exponential distribution}), and further multiplied by a term that describes whether the short segment repels or attracts the nearby trajectories. The last term thus has to take the form $\exp(-\partial_xv\Delta t)$, where $\partial_xv=\partial_x^2S/m$ is the rate of attraction or repulsion of the nearby trajectories when the sign is negative or positive respectively. One therefore has    
\begin{eqnarray}
\rho(x+\Delta x,t+\Delta t)\sim \rho(x,t)e^{-\frac{2}{\hbar}|\Delta S-\Delta S_c|-\frac{\partial_x^2S}{m}\Delta t}. 
\label{probability density anzat}
\end{eqnarray}  
Note that Eq. (\ref{probability density anzat}) does not differentiate between two different cases of $\Delta S\ge\Delta S_c$ and $\Delta S <\Delta S_c$. 

Next, assuming that $\Delta x$ and $\Delta t$ are sufficiently small, expanding the exponential on the right hand side of Eq. (\ref{probability density anzat}) up to the first order one has 
\begin{eqnarray}
\Delta\rho(x,t)=-\Big[\frac{2}{\hbar}\Big|\Delta S-\Delta S_c\Big|+\frac{1}{m}\partial_x^2S\Delta t\Big]\rho(x,t).
\label{fundamental equation}
\end{eqnarray} 
Further, expanding $\Delta \rho$ and $\Delta S$ as $\Delta f=\frac{\partial f}{\partial t}\Delta t+\frac{\partial f}{\partial x}\Delta x$, and comparing term by term one finally obtains 
\begin{eqnarray}
\frac{\hbar}{2}\frac{\partial_x\rho}{\rho}=\pm\big(\partial_xS_c(x,t)-\partial_xS(x,t)\big),\hspace{8mm}\nonumber\\
\frac{\hbar}{2}\frac{\partial_t\rho}{\rho}=\pm\big(\partial_tS_c(x,t)-\partial_tS(x,t)\big)-\frac{\hbar}{2m}\partial_x^2S. 
\label{fundamental equation}
\end{eqnarray}
Here the ``$+$'' and ``$-$'' signs correspond to the case when $\Delta S\ge \Delta S_c$ and $\Delta S<\Delta S_c$, respectively. It is thus clear that in the regime where the terms containing $\hbar$ in Eq. (\ref{fundamental equation}) are negligible, then the momentum and energy are effectively given by their classical mechanics values: 
\begin{equation}
p=\partial_xS\approx \partial_xS_c,\hspace{2mm}H=-\partial_tS\approx -\partial_tS_c. 
\end{equation}
As expected, formally, classical mechanics (Hamilton's principle) is regained in the limit $\hbar\rightarrow 0$. 

Before proceeding, let us remark that Eq. (\ref{fundamental equation}) can not be interpreted as causal dynamical relation for single event. The left hand side is determined by probability density $\rho(x,t)$ which gives the relative frequency that the particle is at $x$ at time $t$ in infinitely many trials, and the right hand side are dynamical quantities which refer to single event. Equation (\ref{fundamental equation}) is thus descriptive rather than explaining causal relation. Further, to verify the above equations, one has to run in principle infinitely many trajectories. As explicitly shown in the original equation of (\ref{probability density anzat}), Eq. (\ref{fundamental equation}) has to be interpreted as equation for $\rho(x,t)$ given the dynamical variable $S(x,t)$ as in classical physics. Moreover, any causal conclusion about the dynamics of the particle based on Eq. (\ref{fundamental equation}) can only be taken through averaging over the ensemble described by $\rho(x,t)$.

Now let us discuss how the above statistical modification of Hamilton's principle changes the dynamical equation of classical mechanics for ensemble of trajectories. Recall that the latter is given by a pair of coupled equations (\ref{Hamilton-Jacobi equation}) and (\ref{classical continuity equation}). Moreover, notice that the coupling between $\rho(x,t)$ and $S_c(x,t)$ in those equations are one sided. Namely, while $\rho(x,t)$ depends on $S_c(x,t)$ through Eq. (\ref{classical continuity equation}), $S_c(x,t)$ does not depend on $\rho(x,t)$. We shall show that when the Hamilton's principle is modified statistically as in Eq. (\ref{postulate of exponential distribution}), then the coupling between the action and the probability density will be two sided.  

To do this, first, we shall discuss the case of ``$+$'' and ``$-$'' signs in Eq. (\ref{fundamental equation}) separately and combine the resulting equations afterward with additional physical assumption. 

Let us consider the case when $\Delta S\ge\Delta S_c$ so that Eq. (\ref{fundamental equation}) takes ``$+$'' sign. Inserting the upper equation of (\ref{fundamental equation}) into the continuity equation of (\ref{classical continuity equation}) one obtains 
\begin{equation}
\partial_t\rho=-\frac{\hbar}{2m}\partial_x^2\rho-\partial_x\Big(\rho\frac{\partial_xS}{m}\Big).  
\label{Fokker-Planck equation 1}
\end{equation}
On the other hand, substituting both equations in (\ref{fundamental equation}) into the Hamilton-Jacobi equation of (\ref{Hamilton-Jacobi equation}) one gets
\begin{eqnarray}
H=-\partial_tS=\frac{(\partial_xS)^2}{2m}+V-\frac{\hbar^2}{2m}\frac{\partial_x^2R}{R}\hspace{0mm}\nonumber\\
+\frac{\hbar}{2\rho}\Big(\partial_t\rho+\frac{\hbar}{2m}\partial_x^2\rho+\partial_x\Big(\rho\frac{\partial_xS}{m}\Big)\Big),
\label{H-J-A equation decomposed 1}
\end{eqnarray}
where $R\equiv\sqrt{\rho}$ and we have used the following identity
\begin{equation}
\frac{\hbar^2}{8m}\Big(\frac{\partial_x\rho}{\rho}\Big)^2=-\frac{\hbar^2}{2m}\frac{\partial_x^2R}{R}+\frac{\hbar^2}{4m}\frac{\partial_x^2\rho}{\rho}, 
\label{fluctuations decomposition}
\end{equation}
Finally, inserting Eq. (\ref{Fokker-Planck equation 1}) into Eq. (\ref{H-J-A equation decomposed 1}) one obtains
\begin{eqnarray}
H=-\partial_tS=\frac{(\partial_xS)^2}{2m}+V-\frac{\hbar^2}{2m}\frac{\partial_x^2R}{R}. 
\label{Madelung equation}
\end{eqnarray}

Next, let us consider the case when $\Delta S<\Delta S_c$ so that Eq. (\ref{fundamental equation}) takes ``$-$'' sign. Again, inserting the upper equation of (\ref{fundamental equation}) into the classical continuity equation of (\ref{classical continuity equation}) one gets 
\begin{equation}
\partial_t\rho=\frac{\hbar}{2m}\partial_x^2\rho-\partial_x\Big(\rho\frac{\partial_xS}{m}\Big).  
\label{Fokker-Planck equation 2}
\end{equation}  
Further, substituting both equations in (\ref{fundamental equation}) into 
the Hamilton-Jacobi equation of (\ref{Hamilton-Jacobi equation}) one gets
\begin{eqnarray}
H=-\partial_tS=\frac{(\partial_xS)^2}{2m}+V-\frac{\hbar^2}{2m}\frac{\partial_x^2R}{R}\hspace{0mm}\nonumber\\
-\frac{\hbar}{2\rho}\Big(\partial_t\rho-\frac{\hbar}{2m}\partial_x^2\rho+\partial_x\Big(\rho\frac{\partial_xS}{m}\Big)\Big),
\label{H-J-A equation decomposed 2}
\end{eqnarray}
where we have used again Eq. (\ref{fluctuations decomposition}). Substituting Eq. (\ref{Fokker-Planck equation 2}) into Eq. (\ref{H-J-A equation decomposed 2}), one finally obtains
\begin{eqnarray}
H=-\partial_tS=\frac{(\partial_xS)^2}{2m}+V-\frac{\hbar^2}{2m}\frac{\partial_x^2R}{R},\nonumber
\end{eqnarray}
which is exactly equal to Eq. (\ref{Madelung equation}).

We have thus two pairs of equations, one is given by Eqs. (\ref{Fokker-Planck equation 1}) and (\ref{Madelung equation}) if $\Delta S\ge \Delta S_c$, and another one is given by Eqs. (\ref{Fokker-Planck equation 2}) and (\ref{Madelung equation}) if $\Delta S<\Delta S_c$. In both cases, the energy, which is equal to the temporal change of the action $H=-\partial_tS$, is given by Eq. (\ref{Madelung equation}); whereas the temporal change of probability density, $\partial_t\rho$, differs only on the sign of the first term on the right hand side. Moreover, notice that so far we only assume the statistical distribution of the deviations $|\Delta S-\Delta S_c|$ of the action. Hence there is still ambiguity in the choice of the distribution of cases of $\Delta S\ge \Delta S_c$ and $\Delta S<\Delta S_c$, namely the distribution of the ``$\pm$'' signs in Eq. (\ref{fundamental equation}). Now let us proceed to assume that the two cases of dynamics occur equally probably independent of the value of the deviation. Hence, the relative frequency that $\Delta S\ge \Delta S_c$ is equal to the relative frequency that $\Delta S<\Delta S_c$, namely $\mathcal{P}(+)=\mathcal{P}(-)=1/2$ regardless the value of $|\Delta S-\Delta S_c|$. Averaging over this fluctuation, the first term on the right hand side of Eqs. (\ref{Fokker-Planck equation 1}) and (\ref{Fokker-Planck equation 2}) cancel to each other to give 
\begin{equation}
\partial_t\rho=-\partial_x\Big(\rho\frac{\partial_xS}{m}\Big). 
\label{quantum continuity equation}
\end{equation}

Finally Eqs. (\ref{Madelung equation}) and (\ref{quantum continuity equation}) can be recast into a compact equation for complex-valued function $\psi\equiv\sqrt{\rho}\exp(iS/\hbar)$ as \cite{Madelung paper} 
\begin{equation}
i\hbar\partial_t\psi=-\frac{\hbar^2}{2m}\partial_x^2\psi+V\psi.
\label{Schroedinger equation}
\end{equation}
The above equation is just the Schr\"odinger equation if we identify $\hbar=h/(2\pi)$ where $h$ is Planck constant. The Born's rule is evident $|\psi|^2=\rho$. Further, the ensemble average of energy at any moment is given by 
\begin{eqnarray}
\int dx H(x,t)\rho(x,t)=\int dx\Big(\frac{(\partial_xS)^2}{2m}+V-\frac{\hbar^2}{2m}\frac{\partial_x^2R}{R}\Big)\rho\nonumber\\
=\int dx\psi^{*}\Big(-\frac{\hbar^2}{2m}\partial_x^2+V\Big)\psi.\hspace{20mm} 
\label{average energy}
\end{eqnarray} 
The last line is just the quantum mechanical average energy which is conserved by the Schr\"odinger equation of (\ref{Schroedinger equation}) if $V$ is independent of time.  

To conclude, we have discussed the dynamics of ensemble of trajectories of single particle by assuming that the Hamilton's principle of stationary action is not exact. We then postulate that there is a universal short time interval so that the deviation of the action from the stationary action is distributed along the trajectory according to exponential law with average $\hbar/2$. The dynamics of the ensemble of trajectories is then shown to be governed by the Schr\"odinger equation with Born interpretation of wave function if the two opposite signs of the deviation occur equally probably. 

Let us compare the dynamics of ensemble of trajectories developed in this paper with Nelson stochastic dynamics \cite{Nelson stochastic mechanics} and de Broglie-Bohm pilot-wave theory \cite{Bohm-Hiley book}, two dynamical theories for ensemble of trajectories which are also governed by the Schr\"odinger equation. In stochastic dynamics, the particle is assumed to interact with some universal background field so that it undergoes a stochastic Brownian-like trajectory with diffusion constant $\hbar/(2m)$. Hence the dynamics is stochastic. In contrast to this, the dynamics in our model is deterministic. Moreover, in stochastic dynamics, it is the trajectory which is subject to fluctuations, whereas in our dynamical model, it is the action itself which fluctuates.  

On the other hand, similar to our dynamical model, the pilot-wave theory is deterministic. However, in contrast to pilot-wave theory in which the wave function is assumed to be physically real, in our dynamical model, the wave function is merely an artificial mathematical tool to describe the ensemble of trajectories of a single particle. Hence, borrowing the language of pilot-wave theory, in our model, the particle behaves as  if it is guided by a virtual wave, $\psi(x,t)$, satisfying the Schr\"odinger equation of (\ref{Schroedinger equation}). This leads us to expect that similar to the pilot-wave theory, our model of dynamics will also show statistical wave-like pattern in double slit experiment \cite{Philippidis}.

\begin{acknowledgments} 

\end{acknowledgments}


\begin{thebibliography}{10} 

\bibitem{Landau book} L. D. Landau and E. M. Lifshitz, Mechanics, Course of Theoretical Physics,  Butterworth-Heinenann, UK, 1976.    

\bibitem{book on probability} Athanasios Papoulis, Probability, Random Variables, and Stochastic processes, McGraw-Hill, Boston, 1991.     

\bibitem{Madelung paper} E. Madelung, Zeits. F. Phys. 40 (1927) 322.  

\bibitem{Nelson stochastic mechanics} Edward Nelson, Phys. Rev. 150 (1966) 1079; Edward Nelson, Quantum Fluctuations, Princeton University Press, Princeton, 1985.  

\bibitem{Bohm-Hiley book} D. Bohm, Phys. Rev. 85 (1952) 166; D. Bohm and B. Hiley, The Undivided Universe: An ontological interpretation of quantum theory, Routledge, London, 1993. 

\bibitem{Philippidis} C. Philippidis, C. Dewdney and B. J. Hiley, Nuovo Cimento 52 B (1979) 15. 

\end{thebibliography}
\end{document}